\documentclass[a4paper,11pt]{article}
\pdfoutput=1 

\usepackage{jheppub}

\usepackage[T1]{fontenc} 

\usepackage{orcidlink}
\usepackage[compat=1.1.0]{tikz-feynman}
\tikzfeynmanset{warn luatex=false}
\usepackage{mathtools}
\usepackage{enumitem}

\newcommand{\fourPoint}{ {
\begin{tikzpicture}[baseline=0]
\begin{feynman}
\vertex (mid1) at (0,0);
\vertex (mid2) at (1,0);
\vertex (a1) at (-0.7,-0.7){$1$};
\vertex (a2) at (-0.7,0.7){$2$};
\vertex (a3) at (1.7,0.7){$3$};
\vertex (a4) at (1.7,-0.7){$4$};
\diagram{
(mid1) -- [ultra thick](mid2),
(mid1) -- [ultra thick](a1),
(mid1) -- [ultra thick](a2),
(mid2) -- [ultra thick](a3),
(mid2) -- [ultra thick](a4),
};
\end{feynman}
\end{tikzpicture}
}
}

\newcommand{\nPoint}{ {
\begin{tikzpicture}[baseline=0]
\begin{feynman}
\vertex (mid1) at (0,0);
\vertex (mid2) at (1,0);
\vertex (a1) at (0,-0.7);
\vertex (l1) at (0,-1){\small{$1$}};
\vertex (a2) at (-0.5,-0.5);
\vertex (l2) at (-0.7,-0.7){\small{$2$}};
\vertex (a3) at (-0.5,0);
\vertex (l3) at (-0.7,0){\small{$3$}};
\vertex (a4) at (-0.5,0.5);
\vertex (l4) at (-0.7,0.7){\small{$...$}};
\vertex (a5) at (0,0.7);
\vertex (l5) at (0,1){\small{$n-1$}};
\vertex (a10) at (1,0.7);
\vertex (l10) at (1,1){\small{$n$}};
\vertex (a9) at (1.5,0.5);
\vertex (l9) at (2.0,0.7){\small{$n+1$}};
\vertex (a8) at (1.5,0);
\vertex (l8) at (1.9,0){\small{$...$}};
\vertex (a7) at (1.5,-0.5);
\vertex (l7) at (2.1,-0.6){\small{$2n-3$}};
\vertex (a6) at (1,-0.7);
\vertex (l6) at (1.1,-1){\small{$2n-2$}};
\diagram{
(mid1) -- [ultra thick](mid2),
(mid1) -- [ultra thick](a1),
(mid1) -- [ultra thick](a2),
(mid1) -- [ultra thick](a3),
(mid1) -- [ultra thick](a4),
(mid1) -- [ultra thick](a5),
(mid2) -- [ultra thick](a6),
(mid2) -- [ultra thick](a7),
(mid2) -- [ultra thick](a8),
(mid2) -- [ultra thick](a9),
(mid2) -- [ultra thick](a10),
};
\end{feynman}
\end{tikzpicture}
}
}

\title{\boldmath Color Relations and Off-Shell Double-Copy for Towers of Theories}

\author{James Mangan\orcidlink{0000-0002-9713-7446},}
\author{Able Martinez\orcidlink{0009-0006-0317-7735},}
\author{and Joseph Thomas\orcidlink{0009-0000-0686-369X}}
\affiliation{Department of Physical and Environmental Sciences, Colorado Mesa University,\\Grand Junction, Colorado 81521, U.S.A.}

\emailAdd{jmangan@coloradomesa.edu}
\emailAdd{aamartinez3@mavs.coloradomesa.edu}
\emailAdd{jthomas3@mavs.coloradomesa.edu}

\abstract{
Only a handful of elementary color building blocks are known to participate in the double copy, most notably the structure constants of gauge theory.
This work extends the double copy to include the infinite family of totally antisymmetric ``structure constants'' with arbitrarily many indices.
We propose and motivate the Jacobi identities relating these $n$-index structure constants.
For each $n$, we provide a derivatively-coupled scalar theory that is color-dual off-shell.
The lowest level in this tower is two-dimensional Zakharov-Mikhailov theory, with one theory for each higher spacetime dimension.
Each member of the tower exhibits a conserved current associated with color-kinematics duality, a soft theorem, an on-shell recursion relation, and classical conformal invariance.
We also provide a tower of modified non-abelian Chern-Simons theories whose fields couple non-linearly through the $n$-index structure constants.
The modified Chern-Simons theories are trivially topological and classically conformal but generally exhibit unusual features like kinetic mixing.
A potentially unphysical gauge condition is identified that would ensure that the tower is color-dual off-shell.
}

\begin{document} 
\maketitle
\flushbottom

\section{Introduction}
\label{sec:intro}

The past several decades have seen substantial progress in our understanding of on-shell quantities in quantum field theory.
While the initial Lagrangian and final scattering amplitude may be remarkably simple, the intermediate Feynman diagram computation is markedly more complex.
Nowhere is this more apparent than in gravity.
The textbook approach using a metric maintains manifest Lorentz covariance but sacrifices unitarity by overcounting physical degrees of freedom.
Restoring unitarity introduces unphysical gauge redundancy that bloats even the simplest $S$-matrix elements.

Remarkably, Bern, Carrasco, and Johansson (BCJ) found that graviton scattering can be greatly simplified by examining the analogous scattering process in a seemingly unrelated gauge theory \cite{Bern:2008qj, Bern:2010ue}.
The relationship between these two theories rests on ``color-kinematics duality,'' meaning that there exists a form of the gauge theory amplitude such that the kinematic numerators obey the same algebraic relations as the color factors.
``Double copying'' the gauge theory amplitude by replacing the color factors with a second instance of the kinematic numerators reproduces the gravity amplitude.
The intuition behind the double copy is that each gauge boson contributes one unit of spin to produce the spin-two graviton.\footnote{In the field-theory limit, the Kawai-Lewellen-Tye relations coincide with the double copy, where the intuition in the former is that sewing two open strings together forms a closed string \cite{Kawai:1985xq}.}
The simplicity of the double copy has made otherwise intractable supergravity calculations feasible and has driven developments in gravitational wave physics \cite{Bern:2009kd, Bern:2007hh, Bern:2012uc, Bern:2017ucb, Bern:2018jmv, Bern:2019nnu, Bern:2019crd}.

The double copy also points to hidden structure, since the theories it relates seem superficially completely unrelated.
For instance, the Einstein-Hilbert action looks nothing like that of Yang-Mills theory, and the connection between the two theories only becomes apparent at the level of on-shell amplitudes.
However, color-dual amplitudes are typically constructed individually via an ansatz method, which provides little clarity regarding the underlying physics.
An off-shell or Lagrangian-level double copy is thus highly desirable when it exists, as it can both offer physical insight into the duality and provide a direct, declarative prescription for relating theories without resorting to an ansatz.

Although the double copy was discovered for gravity and gauge theory, it has since been extended to many other theories including gauge theories with different interactions, theories with varying amounts of supersymmetry, and various effective field theories \cite{Bern:2019prr, Johansson:2017srf, Broedel:2012rc, Chiodaroli:2013upa, Cachazo:2014xea}.
A central theme in the study of the double copy is determining the full scope of the duality.
Frequently, the direction taken has been to explore the landscape of possible kinematic modifications of the double copy, that is, to construct theories from the standard $f^{abc}$ structure constants appearing in Yang-Mills theory while varying the derivative coupling or field content \cite{Johansson:2017srf, Broedel:2012rc, Cachazo:2014xea}.
Another possibility is to change the color constituents of the theory.
Previously, color-kinematics duality has been established for theories involving the generators themselves $T^a_{ij}$, totally symmetric structure constants $d^{abc}$, as well as antisymmetric four-index structure constants $f^{abcd}$ \cite{Johansson:2014zca, Johansson:2015oia, Carrasco:2022jxn, Carrasco:2026hxf, Bargheer:2012gv, Huang:2012wr, Huang:2013kca}.
The present work pursues this second direction by investigating color modifications of the double copy.

Our primary goal is to extend the handful of color objects compatible with the double copy to include the infinite family of totally antisymmetric ``structure constants'' with arbitrarily many indices $f^{abc...}$.
Since the central ingredients of color-kinematics duality are the algebraic (Jacobi) identities relating the color factors, modifying the color structures requires generalizing the algebraic color relations.
After providing a brief review of color-kinematics duality in Section \ref{sec:background}, we propose the generalization of the Jacobi identities for these $n$-index structure constants in Section \ref{sec:color}.
An identity for the Levi-Civita symbol is used to motivate the precise form of the color relations.
Section \ref{sec:CKtheories} presents four color-dual theories.
Each example is in fact an \emph{infinite tower} of \emph{off-shell} color-dual theories with one member for each value of $n\geq3$.\footnote{The tower in Section \ref{sec:modified_CS_CK} is still infinite but it is only compatible with odd $n$.}
The first tower is the $n$-index variant of bi-adjoint scalar theory \cite{Cachazo:2013iea}.
The theories possess ``color'' currents whose conservation is a consequence of manifest color-kinematics duality.
The second example is a tower of derivatively-coupled scalar theories, where the $n=3$ member is Zakharov-Mikhailov theory \cite{Zakharov:1973pp}.
In addition to off-shell color-kinematics duality, each member of this tower is classically conformal and respects a soft theorem that renders the theory on-shell constructible.
One advantage of an off-shell double copy is that it trivially maps the color current from bi-adjoint scalar theory to a conserved current for this derivatively-coupled scalar theory.
The third tower of theories revolves around Chern-Simons theory.
The theories are trivially topological and classically conformal but exhibit unusual features like kinetic mixing.
A gauge condition is identified that would ensure off-shell color-kinematics duality for every entry in the tower, but it is unclear if the gauge is physically attainable.
The fourth tower of examples is a family of pathological vector theories.
This paper opens the door to many potentially fruitful future directions, some of which are presented in Section \ref{sec:conclusions}.
Appendix \ref{sec:f5appendix} discusses five-index structure constants.
Appendix \ref{sec:CSappendix} contains details of the proof that the modified Chern-Simons theory is gauge invariant.
Finally, while the rest of the paper is concerned with \emph{antisymmetric} color structures, Appendix \ref{sec:ZM_appendix} contains a variation of two-dimensional Zakharov-Mikhailov theory (and its Moyal deformation) that incorporates \emph{symmetric} color structures while maintaining off-shell color-kinematics duality.

\section{Background}
\label{sec:background}

To frame the discussion of the higher-rank color structures examined below, we briefly review the traditional adjoint-type double copy \cite{Bern:2010ue, Bern:2008qj, Bern:2019prr}.
Color-kinematics duality is a statement about the off-shell integrands of gauge theories.
Ignoring coupling constants and numerical factors of $2\pi$ and $i$, every $n$-point $L$-loop gauge theory integrand can be written as
\begin{align}
\label{eq:ampNL}
A_n^L = \sum \limits_\Gamma \frac{1}{S_\Gamma} \int d^{LD} \ell ~ \frac{C_\Gamma N_\Gamma}{P_\Gamma},
\end{align}
where $D$ is the spacetime dimension.
Here the sum extends over all graph topologies $\Gamma$, $S_\Gamma$ is the symmetry factor of the graph, $P_\Gamma$ is the collection of propagators, $N_\Gamma$ is the kinematic numerator made up of dot products of momentum and polarization vectors, and $C_\Gamma$ is the color factor constructed from the structure constants $f^{abc}$ associated with the graph.
Even though $A_n^L$ is an on-shell quantity, there is substantial freedom in how it is written.
For example, the color factors obey algebraic relations (Jacobi identities) of the form
\begin{align}
\label{eq:Jacobi}
C_{\Gamma_i}+C_{\Gamma_j}+C_{\Gamma_k}=0
\end{align}
that can be used to move factors between the various terms in \eqref{eq:ampNL}.
Although the amplitudes of any colored theory can be written as \eqref{eq:ampNL}, a special subset of theories including Yang-Mills (YM) theory respect color-kinematics duality.
An amplitude is color-dual if there exists a representation of the amplitude such that the kinematic numerators $N_\Gamma$ obey the same algebraic relations as the color factors $C_\Gamma$.\footnote{Comparison of the color and kinematic factors in this way requires that every graph $\Gamma$ in \eqref{eq:ampNL} be converted to a cubic one.  This can be accomplished by multiplying and dividing terms in the integrand by inverse propagators.}
In particular, in a color-dual theory if a triplet of graphs $\Gamma_i$, $\Gamma_j$, and $\Gamma_k$ obeys \eqref{eq:Jacobi}, then there is a way to rearrange terms such that the kinematic numerators obey the same Jacobi identity,
\begin{align}
\label{eq:NGammaJacobi}
N_{\Gamma_i}+N_{\Gamma_j}+N_{\Gamma_k}=0.
\end{align}
Once a color-dual description is found, the double copy may be performed by replacing $C_\Gamma$ with $N_\Gamma$,
\begin{align}
\label{eq:DC}
C_\Gamma \to N_\Gamma .
\end{align}
The replacement must be performed on the amplitude of some color-dual theory, but the second theory could be different from the first and its amplitude need not be in color-dual form.
If the amplitude of some (other) color-dual gauge theory is
\begin{align}
\bar{A}_n^L = \sum \limits_\Gamma \frac{1}{S_\Gamma} \int d^{LD} \ell ~ \frac{C_\Gamma \bar{N}_\Gamma}{P_\Gamma} ,
\end{align}
then the gravity amplitude is
\begin{align}
\label{eq:MnL}
M_n^L = \sum \limits_\Gamma \frac{1}{S_\Gamma} \int d^{LD} \ell ~ \frac{N_\Gamma \bar{N}_\Gamma}{P_\Gamma}
\end{align}
even if the barred kinematic numerators do not respect \eqref{eq:NGammaJacobi}.
Although this formulation appears to privilege one set of numerators, the double copy is in fact symmetric between the two gauge theories.
However, the apparent asymmetry can be exploited; only one theory needs to be color-dual off-shell to perform a Lagrangian-level double copy.
In the absence of an off-shell color-dual formulation, it is often simplest to make an ansatz for $N_\Gamma$ and impose the algebraic relations inherited by the color factors.
Solving this system of equations can become a non-trivial linear algebra problem, but it remains one of the most efficient methods for computing $M_n^L$.

To provide a concrete example of the double copy, consider the non-linear sigma model (NLSM), which is a color-dual theory alongside YM theory \cite{Cachazo:2014xea}.
Up to a normalization factor, the NLSM four-point tree amplitude can be written in the form of \eqref{eq:ampNL} as
\begin{align}
A_4 = \frac{C_s N_s}{s} + \frac{C_t N_t}{t} +\frac{C_u N_u}{u},
\end{align}
where $s=(p_1+p_2)^2$, $t=(p_2+p_3)^2$, and $u=(p_1+p_3)^2$ are the familiar Mandelstam variables.
The $s$-channel color factor is $C_s = f^{a_1 a_2 c}f^{c a_3 a_4}$ and the kinematic numerator is $N_s = s(t-u)$, where the other channels can be obtained by relabeling.
This is a color-dual form of the numerators because they respect the same two algebraic relations as the color factors.
First, the $s$-channel color factor and numerator are both antisymmetric under exchange of particles 1 and 2 (or 3 and 4).
Second, after imposing on-shell kinematics, the color factors and kinematic numerators respect the same Jacobi identity,
\begin{align}
&C_s+C_t+C_u=0\\
&N_s+N_t+N_u=0 .
\end{align}
The double copy of the NLSM with itself produces the four-point special Galileon amplitude \cite{Nicolis:2008in, Luty:2003vm, Cheung:2014dqa, Cachazo:2014xea},
\begin{align}
M_4 = \frac{N_s^2}{s}+ \frac{N_t^2}{t} + \frac{N_u^2}{u} \propto stu.
\end{align}

The physical necessity of color-kinematics duality becomes apparent when validating the double copy amplitude $M_n^L$ \cite{Bern:2019prr}.
Returning to the discussion of gauge theory and gravity, in order for $M_n^L$ to be the correct gravity amplitude, it should have the right power counting and pole structure, both of which are automatic.
$M_n^L$ should also be invariant under linear diffeomorphisms, that is, the graviton must respect a Ward identity in each of its left and right indices.
The two requisite Ward identities are inherited from the two gauge theories entering the double copy.
However, demonstrating the Ward identity for gauge theory requires invoking the Jacobi identity for the color factors.
This means that, after sending $C_\Gamma \to N_\Gamma$, the kinematic numerators of one gauge theory must respect the Jacobi identity in order to preserve the Ward identity of the \emph{other} gauge theory.
Linearized diffeomorphism invariance together with power counting and locality are enough to guarantee that $M_n^L$ is indeed the correct gravity amplitude \cite{Arkani-Hamed:2016rak}.

What has been described so far is ``on-shell'' color-kinematics duality where theories such as YM theory and the NLSM obey the duality but only after on-shell kinematics are imposed.
However, there is an even more exceptional class of theories that obeys the duality ``off-shell'' or ``manifestly'' at the Lagrangian level.
A non-exhaustive list of such theories includes bi-adjoint scalar (BAS) theory \cite{Cachazo:2013iea}, self-dual YM (SDYM) theory \cite{Monteiro:2011pc, Bjerrum-Bohr:2012kaa}, Chern-Simons (CS) theory \cite{Ben-Shahar:2021zww}, certain formulations of the non-linear sigma model (NLSM) \cite{Cheung:2016prv, Cheung:2021zvb}, ten-dimensional supersymmetric YM (SYM) theory \cite{Ben-Shahar:2021doh}, two-dimensional YM theory \cite{Ben-Shahar:2024dju, Ben-Shahar:2025dci}, $BF$ theory \cite{Ben-Shahar:2024dju, Ben-Shahar:2025dci}, and Zakharov-Mikhailov (ZM) theory \cite{Zakharov:1973pp, Cheung:2022mix}.
At least at the classical level, many of these theories can be organized into the semi-abelian YM theory presented in \cite{Edison:2023ulf}.
These theories are often instructive in that there is some clear physical principle -- like diffeomorphism invariance or a flat gauge connection -- that endows them with color-kinematics duality.
The theories presented in this paper not only respect a generalized version of color-kinematics duality but they do so off-shell.
Unfortunately, the physical reason they are color-dual is more opaque.
The first step in describing these theories is appropriately generalizing the color relations.

\section{Generalized color relations}
\label{sec:color}

The double copy provides a powerful calculational framework that reveals deep structure between seemingly disparate theories.
It is therefore of interest to construct as broad a class of color-dual theories as possible.
While the double copy was originally formulated in terms of purely adjoint color structures, it is natural to expand the space of color-dual theories by considering more general color objects.
The two key features of the adjoint structure constants are, first, they are fully antisymmetric, and second, they carry three indices.
Modifying either one or both of these properties may lead to additional applications of the double copy.
Relaxing the first property to allow for more general index symmetries has proven fruitful, with color-kinematics duality having been established for theories involving $T^a_{ij}$ and the fully symmetric $d^{abc}$ (see \cite{Johansson:2014zca, Johansson:2015oia, Carrasco:2022jxn, Carrasco:2026hxf} among others).
It is also possible to alter the second property of $f^{abc}$ to investigate antisymmetric color objects with more than three indices.
One such example is Bagger-Lambert-Gustavsson (BLG) theory, which can be formulated in terms of a fully antisymmetric four-index object $f^{abcd}$ \cite{Bagger:2007jr, Gustavsson:2007vu, Bargheer:2012gv, Huang:2012wr, Huang:2013kca}.\footnote{Aharony–Bergman–Jafferis–Maldacena (ABJM) theory will be discussed in Section \ref{sec:BLG_times_ZM}.
The structure constants in ABJM carry four indices but are not fully antisymmetric, so ABJM theory modifies both properties of $f^{abc}$ mentioned above.}
In this work, we consider color relations for totally antisymmetric objects with arbitrarily many indices, $f^{abc...}$.
Although no assumptions are made regarding the underlying construction of these color objects, we will still refer to them as ``structure constants.''

From the perspective of the double copy, the only relevant properties of the color constituents are their algebraic relations.
The double copy is insensitive to the detailed construction of the atomic color building blocks, whether they are $f^{abc}$, $f^{abcd}$, $T^a_{ij}$, $d^{abc}$, etc.
Although index symmetries such as $f^{abc} = -f^{bac}$ constitute valid algebraic relations, they are frequently left implicit, so we focus primarily on the bilinear relations including the Jacobi identity and its variants.
The Jacobi relation for $f^{abc}$ and its generalization to $f^{abcd}$ (see \cite{Bagger:2007jr, Gustavsson:2007vu}) are
\begin{align}
\label{eq:1}& f^{a_1 a_2 c}f^{c a_3 a_4} + f^{a_2 a_3 c}f^{c a_1 a_4} + f^{a_3 a_1 c}f^{c a_2 a_4} =0\\
\label{eq:2}& f^{a_1 a_2 a_3 c} f^{c a_4 a_5 a_6} - f^{a_2 a_3 a_4 c} f^{c a_1 a_5 a_6} + f^{a_3 a_4 a_1 c} f^{c a_2 a_5 a_6} - f^{a_4 a_1 a_2 c} f^{c a_3 a_5 a_6} =0.
\end{align}
For the three-index structure constants, the Jacobi relation consists of a cyclic sum over $a_1$, $a_2$, and $a_3$.
For the four-index structure constants, the analogous algebraic relation involves a cyclic sum over $a_1$, $a_2$, $a_3$, and $a_4$ with alternating signs between terms.
These observations suggest that the relevant identity for $n$-index structure constants is that the cyclic sum over $n$ indices vanishes when each term is weighted by the signature of the corresponding permutation,
\begin{align}
\label{eq:colorrel}
\sum \limits_{\sigma \in C_n} \text{sgn}(\sigma) f^{a_{\sigma_1} a_{\sigma_2}... a_{\sigma_{n-1}} c} f^{c a_{\sigma_n} a_{n+1} a_{n+2}...a_{2n-2}}=0,
\end{align}
where $C_n$ denotes the set of cyclic permutations of $\{1,2,...,n\}$, $\sigma_i$ is the $i$th element of the permutation $\sigma$, and $\text{sgn}(\sigma)=\varepsilon^{\sigma_1 \sigma_2...\sigma_n}$ is the signature of the permutation.
In particular, if the structure constants have an even number of indices, then the terms in the sum carry alternating signs, whereas if the structure constants have an odd number of indices, then all of the terms enter with a positive sign.\footnote{The factor of $\text{sgn}(\sigma)$ implies that \eqref{eq:colorrel} holds even when the sum is extended to all permutations.}
For notational convenience, the signed cyclic sum above will be denoted by $\text{sgn cyc}(a_1,a_2,...,a_n)$, so that the color relation can be written as
\begin{align}
f^{a_1 a_2... a_{n-1} c} f^{c a_n a_{n+1} ...a_{2n-2}} + \text{sgn cyc}(a_1,a_2,...,a_n)=0.
\end{align}
Note that \eqref{eq:colorrel} correctly reproduces \eqref{eq:1} as well as \eqref{eq:2} and also trivially holds for antisymmetric two-index structure constants.

The theories in Section \ref{sec:CKtheories} support \eqref{eq:colorrel} a posteriori, but the color relation can be motivated using alternate means.
The Levi-Civita symbol should be an admissible realization of $f^{abc...}$ for several reasons.
First, the Levi-Civita symbol possesses the required index symmetries.
Second, one realization of the familiar three-index structure constants $f^{abc}$ is the $\varepsilon^{abc}$ of $SO(3)$, while the essentially unique choice for the four-index structure constants $f^{abcd}$ is the $\varepsilon^{abcd}$ of $SO(4)$ \cite{2007arXiv0712.1398N, Gauntlett:2008uf, Papadopoulos:2008sk}.\footnote{See Appendix \ref{sec:f5appendix} for an example of five-index structure constants.}
Since $\varepsilon^{abc...}$ is one specific realization of $f^{abc...}$, if \eqref{eq:colorrel} is the appropriate color relation, then it should hold for the Levi-Civita symbol.

The fact that the Levi-Civita symbol satisfies a relation of the form \eqref{eq:colorrel} is presumably well known, but the result is sufficiently elementary that a proof is included here.
After replacing $f^{abc...}$ with $\varepsilon^{abc...}$, the sum in \eqref{eq:colorrel} reads
\begin{align}
\sum \limits_{\sigma \in C_n} \text{sgn}(\sigma) \varepsilon^{a_{\sigma_1} a_{\sigma_2}... a_{\sigma_{n-1}} c} \varepsilon^{c a_{\sigma_n} a_{n+1} a_{n+2}...a_{2n-2}}.
\end{align}
Consider a non-zero term in the sum, provided one exists.
The Levi-Civita symbol on the left, $\varepsilon^{a_{\sigma_1} a_{\sigma_2}... a_{\sigma_{n-1}} c}$, vanishes unless all of its indices are distinct.
Since the indices range from 1 to $n$, there is a bijection between $\{1, 2,...,n\}$ and $\{a_{\sigma_1}, a_{\sigma_2},...,a_{\sigma_{n-1}}, c\}$.
The Levi-Civita symbol on the right, $\varepsilon^{c a_{\sigma_n} a_{n+1} a_{n+2}...a_{2n-2}}$, similarly vanishes if any of its indices are repeated.
Consequently, $c$ and $a_{\sigma_n}$ must be distinct, meaning that $a_{\sigma_n}$ and exactly one other $a_{\sigma_i}$ must coincide.
Thus, the sum vanishes unless there is precisely one repeated index in $\{a_1, a_2,...,a_n\}$, with all other indices taking on distinct values.
Because of the cyclic nature of the sum, we may take one of the repeated indices to be $a_n$ while the other is $a_k$ for some fixed $k$.
In all but two terms of the sum, both $a_n$ and $a_k$ appear on the left Levi-Civita symbol, causing those terms to vanish.
The two non-zero terms in the sum cancel as follows:
\begin{align}
\begin{split}
&~\varepsilon^{a_1 a_2...a_{n-1} c} \varepsilon^{c a_n a_{n+1} ...a_{2n-2}}\\
&\quad +\text{sgn}(k+1,...,n, 1,...,k) \varepsilon^{a_{k+1}... a_{n-1} a_n a_1... a_{k-1}c} \varepsilon^{c a_k a_{n+1}...a_{2n-2}}
\end{split}\\
& = [\varepsilon^{a_1 a_2...a_{n-1} c} + \varepsilon^{k+1...n 1...k} \varepsilon^{a_{k+1}...a_{n-1} a_k a_1...a_{k-1}c}]\varepsilon^{c a_k a_{n+1}...a_{2n-2}}\\
& = [\varepsilon^{a_1 a_2...a_{n-1} c} + \varepsilon^{k+1...n 1...k} (-1)^{k-1}  \varepsilon^{a_{k+1}...a_{n-1} a_1...a_{k-1} a_k c}]\varepsilon^{c a_k a_{n+1}...a_{2n-2}}\\
& = [\varepsilon^{a_1 a_2...a_{n-1} c} + (-1)^{(n-k)k} \varepsilon^{12...n} (-1)^{k-1} (-1)^{(n-k-1)k}  \varepsilon^{a_1a_2...a_{n-1} c}]\varepsilon^{c a_k a_{n+1}...a_{2n-2}}\\
&= [1+(-1)^{2(n-k)k-1}]\varepsilon^{a_1 a_2...a_{n-1} c}\varepsilon^{c a_k a_{n+1}...a_{2n-2}}=0.
\end{align}
In the second line, we have used that $a_n=a_k$ and that $\text{sgn}(...)$ is itself a Levi-Civita symbol.
In the third line, the factor of $(-1)^{k-1}$ is generated by moving $a_k$ right $k-1$ times in the Levi-Civita symbol.
In the fourth line, the order of the indices is canonicalized by first moving $1$ left $n-k$ times, and then moving $2$ left $n-k$ times, etc., while also moving $a_1$ left $n-k-1$ times, and then moving $a_2$ left $n-k-1$ times, etc., introducing factors of $(-1)^{(n-k)k}$ and $(-1)^{(n-k-1)k}$, respectively.
Thus, the Levi-Civita symbol satisfies \eqref{eq:colorrel}, providing evidence that this is the appropriate generalization of the Jacobi identity.

\section{Color-dual theories}
\label{sec:CKtheories}

Having motivated the color relation in \eqref{eq:colorrel}, we move on to presenting four color-dual theories involving the $n$-index structure constants.
The first is a generalization of BAS theory, the second is a generalization of ZM theory, the third is a modified CS theory, and the fourth is a pathological vector theory.
All of these theories exhibit color-kinematics duality at the off-shell or Lagrangian level, trivializing the double copy, at least when each theory is double-copied with itself.

To illustrate how off-shell color-kinematics duality is proven, we will consider the standard three-index structure constants, but the generalization will be immediate.
There are two kinds of color relations, those coming from antisymmetry of the structure constants and those coming from the $n=3$ version of \eqref{eq:colorrel}, i.e., the ordinary Jacobi identity.
The former color relations require that the kinematic vertices are totally antisymmetric, which will be automatic for all of the theories we present.
The non-trivial color relation is the Jacobi identity,
\begin{equation}
\label{eq:C_Jacobi}
C\left[\fourPoint\right] + \text{cyc}(1,2,3) = 0 ,
\end{equation}
where $C[...]$ denotes the color factor of the associated graph.
This is nothing more than a graphical representation of the $n=3$ case of \eqref{eq:colorrel}.
The condition for \emph{on-shell} color-kinematics duality of this four-point process is
\begin{equation}
\label{eq:N_Jacobi}
N\left[\fourPoint\right] + \text{cyc}(1,2,3) = 0 ,
\end{equation}
where $N[...]$ denotes the kinematic numerator of the specified graph.
To prove \emph{off-shell} color-kinematics duality, note that every non-trivial color relation stems from \eqref{eq:C_Jacobi} where \eqref{eq:C_Jacobi} might have been embedded in some larger process.
If \eqref{eq:N_Jacobi} is proved with all legs off-shell, then \eqref{eq:N_Jacobi} can be automatically inserted into any larger process, proving the duality in general.

When a theory involves the $n$-index structure constants rather than the three-index version, the natural framework for color-kinematics duality is $n$-valent graphs instead of cubic graphs.
Proving the duality off-shell means showing that
\begin{equation}
\label{eq:N_Jacobi_general}
N\left[\nPoint\right] + \text{sgn cyc}(1,2,...,n) = 0
\end{equation}
holds off-shell.
This is simply a graphical representation of the kinematic version of \eqref{eq:colorrel}.
In what follows, coupling constants have been suppressed for brevity.

\subsection{Generalized bi-adjoint scalar theory}

Bi-adjoint scalar (BAS) theory,
\begin{align}
\label{eq:BAS}
\mathcal{L} = - \frac{1}{2} \partial_\mu \phi^{a\bar{a}} \partial^\mu \phi^{a\bar{a}} + \frac{1}{3!} f^{abc} \bar{f}^{\bar{a}\bar{b}\bar{c}} \phi^{a\bar{a}} \phi^{b\bar{b}} \phi^{c\bar{c}} ,
\end{align}
holds a privileged position in the adjoint double copy involving three-index structure constants \cite{Cachazo:2013iea}.\footnote{Unsurprisingly, BAS theory also plays a special role in the Cachazo-He-Yuan (CHY) construction and the KLT relations \cite{Cachazo:2013hca, Cachazo:2013iea}.}
Off-shell color-kinematics duality, for example in the language of \eqref{eq:N_Jacobi}, holds trivially because the color factors and kinematic numerators are literally identical after exchanging barred and unbarred indices.
The double copy is implemented via the mechanical replacement rule $f^{abc} \to \bar{f}^{\bar{a}\bar{b}\bar{c}}$, meaning that BAS theory is the identity for the BCJ product.
Color-kinematics duality for BAS theory is associated with color conservation, whose symmetry and Noether current are
\begin{align}
&\phi^{a \bar{a}} \to \phi^{a \bar{a}} + f^{abc} \phi^{b\bar{a}}\theta^c\\
&J_\mu^a = f^{abc} \phi^{b\bar{a}} \partial_\mu \phi^{c\bar{a}},
\end{align}
respectively, where the analogous statement holds for barred color.
Conservation of the color current critically relies on the Jacobi identity as follows.
The divergence of the current produces two terms $\partial^\mu J_\mu^a = f^{abc} \partial^\mu \phi^{b\bar{a}} \partial_\mu \phi^{c\bar{a}} + f^{abc} \partial^\mu \phi^{b\bar{a}} \Box \phi^{c\bar{a}}$.
The first term vanishes by antisymmetry of $f^{abc}$ while the second term vanishes after inserting the equation of motion, relabeling terms, and then applying the Jacobi identity.

The generalization of \eqref{eq:BAS} to the $n$-index structure constants is obvious,
\begin{align}
\label{eq:generalizedBAS}
\mathcal{L} = -\frac{1}{2} \partial_\mu \phi^{a\bar{a}} \partial^\mu \phi^{a\bar{a}} + \frac{1}{n!} f^{a_1 a_2 ... a_n} \bar{f}^{\bar{a}_1 \bar{a}_2 ... \bar{a}_n} \phi^{a_1\bar{a}_1} \phi^{a_2\bar{a}_2} ... \phi^{a_n\bar{a}_n} .
\end{align}
Even though no assumptions are made as to how the fields actually transform, we will refer to this as \emph{generalized BAS theory}.
Color-kinematics duality in the form of \eqref{eq:N_Jacobi_general} is automatically fulfilled off-shell since the kinematic numerators are identical to the associated color factors except with barred and unbarred indices interchanged.
The double copy is implemented via the replacement rule $f^{abcd...} \to \bar{f}^{\bar{a} \bar{b} \bar{c} \bar{d}...}$.
Since \eqref{eq:generalizedBAS} is a scalar theory, it does not require ghosts at loop level, so color-kinematics duality persists to all orders in perturbation theory.

Similar to ordinary cubic BAS theory, color-kinematics duality for \eqref{eq:generalizedBAS} is associated with color conservation
\begin{align}
&\label{eq:generalizedBASsym}
\phi^{a_1 \bar{a}} \to \phi^{a_1 \bar{a}} + f^{a_1...a_n} \phi^{a_2 \bar{a}} \theta^{a_3...a_n}\\
& \label{eq:generalizedBAScurrent}
J_\mu^{a_1...a_{n-2}} = f^{a_1...a_n} \phi^{a_{n-1} \bar{a}} \partial_\mu\phi^{a_n\bar{a}} ,
\end{align}
where the analogous statement holds for barred color.
As before, conservation of the current in \eqref{eq:generalizedBAScurrent} relies on antisymmetry of $f^{a_1...a_n}$ and the generalized Jacobi identity in \eqref{eq:colorrel}.
It is also interesting to consider a ``contracted'' or ``dualized'' current
\begin{align}
\tilde{J}_\mu^{ab} = f^{a b c_1...c_{n-2}} J_\mu^{c_1...c_{n-2}}.
\end{align}
This dualized current has the advantage that it always carries two color indices independent of the number of color indices on the structure constants.
Since this is simply a contraction of the original current, it will be conserved if the original current is.
If the structure constants are a Levi-Civita symbol, the contracted current reduces to the particularly simple combination
\begin{align}
\phi^{a\bar{a}} \partial_\mu \phi^{b\bar{a}} - \phi^{b\bar{a}} \partial_\mu \phi^{a\bar{a}}.
\end{align}
It is not accidental that this is reminiscent of the current for a complex scalar since the latter can be seen as having a rotational symmetry.
The barred and unbarred color structures are completely symmetric, so the barred color transformation and current read
\begin{align}
&\phi^{a \bar{a}_1} \to \phi^{a \bar{a}_1} + \bar{f}^{\,\bar{a}_1...\bar{a}_n} \phi^{a \bar{a}_2} \theta^{\,\bar{a}_3...\bar{a}_n}\\
& \label{eq:generalizedBAScurrent_dual}
\bar{J}_\mu^{\bar{a}_1...\bar{a}_{n-2}} = \bar{f}^{\,\bar{a}_1...\bar{a}_n} \phi^{a \bar{a}_{n-1}} \partial_\mu\phi^{a \bar{a}_n} ,
\end{align}
which we record for later convenience.
Although the barred and unbarred color factors are on completely equal footing, this is not true in the next example.

\subsection{Generalized Zakharov-Mikhailov theory}

A primary result of this work is that the derivatively-coupled scalar theory
\begin{align}
\label{eq:generalizedZM}
\mathcal{L} = -\frac{1}{2} \partial_\mu \phi^a \partial^\mu \phi^a + \frac{1}{(D+1)!}f^{a_1...a_{D+1}} \varepsilon^{\mu_1...\mu_D} \partial_{\mu_1} \phi^{a_1}...\partial_{\mu_D} \phi^{a_D} \phi^{a_{D+1}}
\end{align}
is color-dual off shell.
The spacetime dimension $D$ appears explicitly in the theory because the structure constants carry one more index than the Levi-Civita symbol.
The theory above is a set of scalar theories, one for each spacetime dimension, so we are actually presenting an infinite family of color-dual theories.
As a scalar theory, \eqref{eq:generalizedZM} does not require ghosts at loop level, so once off-shell color-kinematics duality has been demonstrated at tree level, it will hold at every loop order.

Before proving off-shell color-kinematics duality for the theory in \eqref{eq:generalizedZM}, it is useful to understand how this theory relates to other known color-dual theories.
The $D=2$ case of \eqref{eq:generalizedZM} coincides with Zakharov-Mikhailov (ZM) theory, so we will refer to \eqref{eq:generalizedZM} as \emph{generalized Zakharov-Mikhailov (ZM) theory} \cite{Zakharov:1973pp}.
ZM theory is classically equivalent to the NLSM in two dimensions since the (dualized) chiral current shares the same equation of motion as ZM theory.
When the propagators in self-dual YM (SDYM) theory are restricted to two dimensions, the theory reduces to ZM theory.
Thus, the generalized ZM theory in \eqref{eq:generalizedZM}, the original 2D ZM theory, the NLSM, and SDYM theory are all related to each other at least in two dimensions, and all four theories are color-dual off-shell \cite{Cheung:2022mix, Cheung:2016prv, Cheung:2021zvb, Monteiro:2011pc, Bjerrum-Bohr:2012kaa}.
Generalized ZM theory has several other interesting properties, but this section will focus exclusively on off-shell color-kinematics duality with all other observations delayed until Section \ref{sec:BAS_times_ZM}.

At a high level, off-shell color-kinematics duality for generalized ZM theory holds because $D+1$ vectors must be linearly dependent in $D$ dimensions.
To prove the duality, the off-shell kinematic numerators in \eqref{eq:N_Jacobi_general} must be calculated and their sum shown to vanish.
The necessary Feynman rule for the $(D+1)$-point vertex in generalized ZM theory is
\begin{equation}
\label{eq:ZM_vert}
i^{D+1} f^{a_1a_2...a_{D+1}} \varepsilon( p_1, p_2, ..., p_D ) ,
\end{equation}
where the shorthand
\begin{equation}
\varepsilon( X_1, X_2,..., X_D) = \varepsilon_{\mu_1 \mu_2 ... \mu_D} X_1^{\mu_1} X_2^{\mu_2} ... X_D^{\mu_D}
\end{equation}
has been employed.
Once momentum conservation is taken into account, $\varepsilon( p_1, p_2,..., p_D )$ is totally antisymmetric in all $D+1$ (not just $D$) particles entering the vertex, paralleling the color factor.\footnote{Since the kinematic vertex is totally antisymmetric, \eqref{eq:ZM_vert} is Bose symmetric as written without the need for any additional permutation sum.}
The kinematic numerators in \eqref{eq:N_Jacobi_general} are generated by gluing together two vertices and then stripping off the structure constants and propagators.
Up to an irrelevant overall factor, \eqref{eq:N_Jacobi_general} becomes
\begin{equation}
\label{eq:KinJacIdTemp}
\varepsilon( p_1,...,p_D ) \varepsilon( p_{D+1},...,p_{2D} ) + \text{sgn cyc}(1,2,...,D+1) = 0 .
\end{equation}
Since the momentum vectors $p^\mu_{D+2}, ...,p^\mu_{2D}$ never cycle in the sum, it is enough to show that
\begin{align}
\label{eq:pVecJacobiId}
0 &= \varepsilon( p_1, p_2,..., p_D ) p_{D+1}^\mu  + \text{sgn cyc}(1,2,..., D+1) \\
\label{eq:pVecJacobiIdExpanded}
&= \varepsilon( p_1, p_2,..., p_D ) p_{D+1}^\mu + ... +(-1)^{Di} \varepsilon(p_{i+1}, ..., p_{D+1}, p_1, ..., p_{i-1} ) p_i^\mu + ...,
\end{align}
where \eqref{eq:KinJacIdTemp} follows by dotting $\varepsilon_{\mu \mu_{D+2}... \mu_{2D}} p_{D+2}^{\mu_{D+2}}... p_{2D}^{\mu_{2D}}$ into \eqref{eq:pVecJacobiId}.
The signed cyclic sum has been written out in the second line for later convenience.
Equation \eqref{eq:pVecJacobiId} is essentially a $D$-dimensional variant of Schouten's identity.
The equation can also be interpreted in terms of Cramer's rule as the solution to $P \mathbf{x} = \det(P) \mathbf{p}_{D+1}$ where the momentum vectors $\mathbf{p}_1$ to $\mathbf{p}_D$ form the columns of the matrix $P$.

To prove \eqref{eq:pVecJacobiIdExpanded}, note that the $D+1$ momentum vectors must be linearly dependent, so there is a non-trivial linear relation of the form $\beta_1 p_1^\mu+\beta_2 p_2^\mu+...=0$.
Without loss of generality, the coefficient of $p_{D+1}^\mu$ may be taken to be non-zero, in which case the linear relation may be rearranged to obtain
\begin{equation}
p_{D+1}^\mu = b_1 p_1^\mu + ... b_D p_D^\mu .
\end{equation}
Inserting this into the first term in \eqref{eq:pVecJacobiIdExpanded} produces
\begin{equation}
\label{eq:JacCancelMe}
\varepsilon( p_1, p_2,..., p_D ) (b_1 p_1^\mu + ... b_D p_D^\mu) ,
\end{equation}
while the remaining terms in \eqref{eq:pVecJacobiIdExpanded} take the form
\begin{align}
&(-1)^{Di} \varepsilon( p_{i+1}, ..., p_D, p_{D+1}, p_1, ..., p_{i-1} ) p_i^\mu\\
=&(-1)^{Di} b_i \varepsilon( p_{i+1}, ...,p_D, p_i, p_1,...,p_{i-1} ) p_i^\mu .
\label{eq:AngleMinuses}
\end{align}
Canonicalizing the order of the vectors in $\varepsilon(...)$ introduces several minus signs.
First, $p_i$ is shifted to the far left in the Levi-Civita contraction, producing $D-i$ minus signs.
Next, $p_1$ is moved $D-i+1$ positions to the left, then $p_2$ is also moved $D-i+1$ positions to the left, etc., for a total of $(D-i+1)(i-1)$ minus signs.
In the end, \eqref{eq:AngleMinuses} becomes
\begin{align}
&(-1)^{Di + (D-i)+(D-i+1)(i-1)} b_i \varepsilon( p_1, p_2,..., p_D ) p_i^\mu\\
=&-b_i \varepsilon( p_1, p_2,..., p_D ) p_i^\mu ,
\end{align}
which exactly cancels \eqref{eq:JacCancelMe}.
Thus \eqref{eq:pVecJacobiIdExpanded} vanishes, proving off-shell color-kinematics duality for generalized ZM theory.

Since color-kinematics holds off-shell, the double copy can be implemented via the simple replacement rule
\begin{align}
\label{eq:DC_rule}
f^{a_1 ... a_{D+1}} X_1^{a_1} ...X_{D+1}^{a_{D+1}} &\to \varepsilon^{\mu_1...\mu_D} \partial_{\mu_1} X_1 ... \partial_{\mu_D} X_D X_{D+1} ,
\end{align}
where the $X_i$ fields are generic objects that can carry additional indices.
The left hand side of \eqref{eq:DC_rule} is totally antisymmetric in the $X_i$ objects but the right hand side is only totally antisymmetric after taking into account integration by parts or equivalently total momentum conservation.
An explicit permutation sum on the right hand side can be included if necessary.
While color-kinematics duality is often associated with volume preserving diffeomorphisms, the connection is less obvious here because the left hand side of \eqref{eq:DC_rule} does not originate from a commutator \cite{Monteiro:2011pc, Bjerrum-Bohr:2012kaa, Ben-Shahar:2021zww, Cheung:2016prv, Cheung:2021zvb, Cheung:2022mix}.
The remainder of this subsection is devoted to three double copies with generalized ZM theory.

\subsubsection{BAS $\otimes$ ZM}
\label{sec:BAS_times_ZM}

Applying the double-copy replacement rule in \eqref{eq:DC_rule} to the barred color indices of the generalized BAS theory in \eqref{eq:generalizedBAS} yields \eqref{eq:generalizedZM}, as required.
As this is merely a consistency check, it provides an opportunity to elaborate on the properties of generalized ZM theory besides manifest color-kinematics duality.

An important question is whether generalized ZM theory possesses non-trivial amplitudes for $D\geq3$.
This concern stems from the fact that the amplitudes of ordinary 2D ZM theory are known to be quite subtle.
The intricacies of the two-dimensional amplitudes are linked to the theory's remarkable properties like classical integrability and a lack of particle production \cite{Zakharov:1973pp, Polyakov:1980ca, Curtright:1994be, Gabai:2018tmm, Nappi:1979ig, Hoare:2018jim}.
Many of the subtleties originate from the massless on-shell conditions in two dimensions, where similar issues would not be expected in higher dimensions.
In two dimensions, every on-shell particle is either a left-mover or a right-mover, meaning that propagators involving only left-movers (or right-movers) exhibit collinear singularities.
As a result, even tree amplitudes must be carefully regulated.
Furthermore, Gram determinant identities proliferate in two dimensions, so it is not uncommon for both the numerator and denominator of an amplitude to vanish simultaneously.
After careful consideration, the color-ordered amplitudes with alternating left- and right-movers vanish \cite{Gabai:2018tmm}.\footnote{This vanishing property was shown for the 2D NLSM but extends to 2D ZM theory since the two theories are equivalent at tree level.}

Given the complexity of ZM theory in two dimensions, it is desirable to determine whether the amplitudes in higher dimensions generally vanish.
The simplest generalized ZM amplitude beyond two dimensions is the three-dimensional six-point tree amplitude,
\begin{equation}
\label{eq:ZM_3D_6pt_amp}
A_6 = f^{a_1 a_2 a_3 c} f^{c a_4 a_5 a_6} \frac{\varepsilon( p_1, p_2, p_3) \varepsilon(p_4, p_5, p_6)  }{s_{123}} + \text{perm}(1,2,3,4,5,6) ,
\end{equation}
where $s_{123} = (p_1+p_2+p_3)^2$ and the sum extends over all permutations of external legs.
A priori this amplitude could vanish due to some hidden Gram determinant identity or the four-index Jacobi identity in \eqref{eq:colorrel}.
However, we have explicitly checked that this amplitude is non-zero.
This can be accomplished by setting $f^{abcd}$ to be $\varepsilon^{abcd}$ and then choosing particular values for the color indices.
Exact numerical on-shell kinematics can be generated by picking random Pythagorean triples for the momentum vectors and then separately scaling each vector to enforce total momentum conservation.\footnote{In $D$ dimensions, momentum vectors with rational components can be constructed by going to light-cone coordinates, $p_\pm=(p_0 \pm p_1)/\sqrt{2}$ and $p_\perp = \{p_2,p_3,...\}$.  If $p_-$ is a rational multiple of $\sqrt{2}$ and $p_\perp$ has rational entries, then setting $p_+ = p_\perp^2/2p_-$ results in a null vector $p^\mu$ with purely rational components.}
The result is that $A_6$ is unambiguously non-zero for generic kinematics, suggesting that the generalized ZM theory in \eqref{eq:generalizedZM} is a fully interacting theory for $D\geq3$.

By virtue of the off-shell double-copy, generalized ZM theory inherits two conserved currents from generalized BAS theory.
Applying the replacement rule in \eqref{eq:DC_rule} to \eqref{eq:generalizedBAScurrent} produces a conserved color current\footnote{Recall that \eqref{eq:generalizedZM} relates the spacetime dimension to the rank of the structure constants through $n=D+1$.}
\begin{align}
J_\mu^{a_1...a_{D-1}} = f^{a_1...a_{D+1}} \phi^{a_{D}} \partial_\mu\phi^{a_{D+1}} .
\end{align}
Conservation of this current proceeds exactly as before.
However, it is less obvious how to apply the replacement rule in \eqref{eq:DC_rule} to \eqref{eq:generalizedBAScurrent_dual} because not all of the barred color indices are contracted.
One natural option is to replace uncontracted color indices with spatial ones to produce
\begin{align}
\label{eq:ZM_kin_current}
J_\mu^{\mu_1...\mu_{D-2}} = \varepsilon^{\mu_1...\mu_D} \partial_{\mu_{D-1}} \phi^a \partial_\mu \partial_{\mu_D} \phi^a .
\end{align}
Conservation of this kinematic current can be seen as follows.
Since the goal is to show that the current is divergenceless, overall numerical factors will be omitted.
Taking the divergence of the current, inserting the equation of motion, and rearranging terms yields
\begin{align}
\partial^\mu J_\mu^{\mu_1...\mu_{D-2}} \sim \varepsilon^{\mu_1...\mu_D} \varepsilon^{\nu_1...\nu_D} f^{a_1...a_{D+1}} \partial_{\mu_{D-1}} \phi^{a_{D+1}} \partial_{\mu_D} \left( \partial_{\nu_1} \phi^{a_1}...\partial_{\nu_D} \phi^{a_D} \right) .
\end{align}
The product of Levi-Civita symbols can be converted to a sum of metrics with alternating signs that enforces the antisymmetry of the indices $\nu_1$,...,$\nu_D$.
The antisymmetry in $\nu_1$,...,$\nu_D$ is already enforced via the structure constants $f^{a_1...a_{D+1}}$, so in this particular instance we may take $\varepsilon^{\mu_1...\mu_D} \varepsilon^{\nu_1...\nu_D} \to \eta^{\mu_1 \nu_1} ...\eta^{\mu_D \nu_D}$ which produces
\begin{align}
\partial^\mu J_\mu^{\mu_1...\mu_{D-2}} & \sim f^{a_1...a_{D+1}} \partial_{\mu_{D-1}} \phi^{a_{D+1}} \partial_{\mu_D} \left( \partial^{\mu_1} \phi^{a_1}...\partial^{\mu_D} \phi^{a_D} \right) \\
& \begin{aligned}
\mathllap{\phantom{ \partial^\mu J_\mu^{\mu_1...\mu_{D-2}} }} & \sim f^{a_1...a_{D+1}} \partial_{\mu_{D-1}} \phi^{a_{D+1}} \\
& \quad \times [ \partial_{\mu_D} (\partial^{\mu_1} \phi^{a_1} ... \partial^{\mu_{D-2}} \phi^{a_{D-2}} \partial^{\mu_D} \phi^{a_D}) \partial^{\mu_{D-1}} \phi^{a_{D-1}} \\
& \quad + (\partial^{\mu_1} \phi^{a_1} ... \partial^{\mu_{D-2}} \phi^{a_{D-2}} \partial^{\mu_D} \phi^{a_D}) \partial_{\mu_D} \partial^{\mu_{D-1}} \phi^{a_{D-1}} ] .
 \end{aligned}
\end{align}
The derivative was carefully distributed in the last line to strategically group terms based on index symmetries.
The first term in the square brackets vanishes after exchanging $a_{D+1}$ and $a_{D-1}$, while the second term in the square brackets vanishes after swapping $\mu_D$ with $\mu_{D-1}$ and $a_{D+1}$ with $a_D$.
Thus the current is conserved.
A kinematic current for the NLSM was presented in \cite{Cheung:2021zvb} that was similarly obtained through an off-shell double copy of a BAS current.
In \cite{Cheung:2021zvb} it was argued that the NLSM current is actually the derivative of yet another conserved quantity and as such it is not associated with a symmetry.
The kinematic current for generalized ZM theory may share the same shortcomings.

The next noteworthy property of generalized ZM theory is that it is conformal, at least classically, like many of the other theories appearing in the double copy \cite{Loebbert:2018xce, Cheung:2020qxc, Farnsworth:2021ycg}.
The Lagrangian in \eqref{eq:generalizedZM} is a linear combination of a massless scalar kinetic term and a derivatively-coupled interaction.
Like a free scalar, the kinetic term is conformal in any dimension when supplemented with the appropriate improvement terms.
The interaction term only depends on the Levi-Civita symbol so it does not contribute to the energy-momentum tensor, $T^{\mu\nu} = 2 \tfrac{\delta S}{\delta g_{\mu\nu}}$.
Conformal symmetry is likely anomalous since the coupling constant is dimensionful for $D\geq 3$.
Even in two dimensions, the beta function is non-zero for generic color groups \cite{Nappi:1979ig}.

Before moving on to double copying generalized ZM theory with itself, \eqref{eq:generalizedZM} is special in yet another way.
Generalized ZM theory has a soft theorem that renders the theory on-shell constructible.
In the limit where $p_i$ is soft, the $n$-point tree amplitude $A_n$ of generalized ZM theory vanishes as $\mathcal{O}(p_i)$.
This is an exact parallel of the Adler zero of the NLSM, but the reasoning here is simpler \cite{Adler:1964um}.
Every diagram contributing to $A_n$ has at least one vertex \eqref{eq:ZM_vert} that scales as $\mathcal{O}(p_i)$, so the entire amplitude should vanish at least as fast as $p_i$.
The only obvious way to spoil this behavior would be if a propagator became singular in the soft limit.
This is impossible for generic kinematics when $D\geq 3$ because each propagator involves the sum of at least three external momenta.\footnote{We have explicitly verified that \eqref{eq:ZM_3D_6pt_amp} has the expected soft behavior.}
This soft theorem in turn means that generalized ZM theory is on-shell constructible \cite{Cheung:2015ota}.
As opposed to the two-line shift of \cite{Britto:2005fq}, the method in \cite{Cheung:2015ota} is to extend the amplitude to a complex function of $z$ by shifting $D+1$ momenta.
As with all on-shell recursion relations, the linchpin is the behavior of the amplitude as $z$ approaches infinity.
The technique in \cite{Cheung:2015ota} is to multiply the amplitude by a known prefactor that leverages the soft information of the theory to improve the large-$z$ fall off.
The theory is recursively constructible provided that $m<\sigma n$, where the interactions of the theory take the form $\partial^m\phi^n$ and the amplitude scales as $p_i^\sigma$ in the soft limit.
For the case of generalized ZM theory, $\sigma=1$, $m=D$, and $n=D+1$ so the requisite condition is fulfilled.

\subsubsection{ZM $\otimes$ ZM}

To double copy generalized ZM theory with itself, the replacement rule in \eqref{eq:DC_rule} is applied to \eqref{eq:generalizedZM} to produce
\begin{align}
\label{eq:generalizedSG}
\mathcal{L} &= -\frac{1}{2} \partial_\mu \phi \partial^\mu \phi + \frac{1}{(D+1)!}  \varepsilon^{\mu_1...\mu_D} \varepsilon^{\nu_1...\nu_D} \phi \partial_{\mu_1} \partial_{\nu_1} \phi...\partial_{\mu_D} \partial_{\nu_D} \phi\\
&= -\frac{1}{2} (\partial \phi)^2 + \frac{1}{D+1} \phi \det(\partial_\mu \partial_\nu\phi) ,
\end{align}
which describes a Galileon-type theory \cite{Nicolis:2008in, Luty:2003vm}.
This theory picks out two specific terms in the $D$-dimensional Galileon effective field theory, namely, the term with the fewest fields and the term with the most fields.
In two dimensions, \eqref{eq:generalizedSG} agrees with the square of ordinary 2D ZM theory as expected \cite{Cheung:2022mix}.
The scattering amplitudes experience the usual subtleties in two dimensions, but the theory turns out to be free \cite{deRham:2013hsa, DeRham:2014wnv, Millington:2017sea, Kampf:2014rka}.
Just like with generalized ZM theory, it is natural to wonder whether the amplitudes are non-zero in higher dimensions.
In three dimensions, the simplest non-trivial scattering amplitude is the six-point tree amplitude,
\begin{equation}
A_6 = \frac{\varepsilon( p_1, p_2, p_3)^2 \varepsilon( p_4, p_5, p_6)^2}{s_{123}} + \text{perm}(1,2,3,4,5,6) ,
\end{equation}
which we have explicitly verified is generically non-zero.
The three-dimensional version of \eqref{eq:generalizedSG} is actually none other than the \emph{special} Galileon theory restricted to three dimensions \cite{Cachazo:2014xea, Cheung:2014dqa, Hinterbichler:2015pqa, Kampf:2014rka}.
Note that while special Galileons are typically obtained by double copying the NLSM with itself, \eqref{eq:generalizedSG} was obtained by double copying the three-dimensional version of \eqref{eq:generalizedZM} with itself, which has entirely different power counting and color structure from the NLSM \cite{Cachazo:2014xea}.

Based on the power counting of the vertex, \eqref{eq:generalizedSG} has a $\mathcal{O}(p_i^2)$ soft theorem.\footnote{The two-dimensional case is subtle, but it is impossible for generic kinematics in higher dimensions to cause a propagator to vanish, spoiling the naive soft behavior.}
In $D=3$ the soft behavior gets enhanced to $\mathcal{O}(p_i^3)$ since the theory coincides with special Galileons, but we have checked that the four-dimensional eight-point tree amplitude vanishes as $\mathcal{O}(p_i^2)$ as naively expected.
The soft theorem renders the theory in \eqref{eq:generalizedSG} on-shell constructible \cite{Cheung:2015ota}.
In the language introduced at the end of Section \ref{sec:BAS_times_ZM}, $m=2D$, $n=D+1$, and $\sigma=2$, which fulfills the on-shell constructibility requirement $m < \sigma n$.
Lastly, the theory in \eqref{eq:generalizedSG} is classically conformal for the same reasons as generalized ZM theory.

\subsubsection{BLG $\otimes$ ZM}
\label{sec:BLG_times_ZM}

Double copying with ZM theory requires a $D$-dimensional color-dual theory with structure constants that have $D+1$ indices.
Bagger-Lambert-Gustavsson (BLG) theory is precisely such a theory \cite{Bagger:2007jr, Gustavsson:2007vu}.
It is a three-dimensional $\mathcal{N}=8$ Chern-Simons theory that describes the interaction of M2 branes and involves four-index structure constants.
Each of the eight directions transverse to the brane ($I=3,4,...,10$) generates a scalar $X_a^I$ valued in the color algebra associated with $f^{abcd}$.
As shown in \cite{Bagger:2007jr}, BLG theory can be constructed by first creating a scalar theory that is invariant under the global symmetry\footnote{Note that one index of the structure constants in \eqref{eq:BLGsym} is treated as special.  This is also the case for the kinematic structure constants in \eqref{eq:DC_rule} as well as the kinematic structure constants in 2D ZM theory, the NLSM, and SDYM theory \cite{Cheung:2022mix, Cheung:2021zvb, Monteiro:2011pc}.  Any ambiguities or subtleties can typically be resolved by explicitly antisymmetrizing the indices of the structure constants.}
\begin{equation}
\label{eq:BLGsym}
X^I_a \to X_a^I + f^{cdb}{}_{a} \theta_{cd} X_b^I .
\end{equation}
This is the same as the ``color rotation'' in \eqref{eq:generalizedBASsym} for the $n=4$ version of generalized BAS theory.
However, that is where the similarity with the generalized BAS theory ends because the interaction term necessary for BLG theory is a sextic,
\begin{equation}
V=\frac{1}{12} f^{abcd} f^{efg}{}_{d} X_a^I X_b^J X_c^K X_e^I X_f^J X_g^K ,
\end{equation}
quite unlike the quartic theory obtained by setting $n=4$ in \eqref{eq:generalizedBAS}.
Gauging the symmetry in the sextic scalar theory produces the complete BLG Lagrangian
\begin{equation}
\label{eq:BLGLagrangian}
\begin{split}
\mathcal{L} =& -\frac{1}{2} D_\mu X^{aI} D^\mu X_a^I -V + \frac{i}{2} \bar{\Psi}^a \Gamma^\mu D_\mu \Psi_a + \frac{i}{4} f^{abcd} \bar{\Psi}_b \Gamma_{IJ} X_c^I X_d^J \Psi_a\\
& + \frac{1}{2} \varepsilon^{\mu\nu\lambda} \left( f^{abcd} A_{\mu ab} \partial_\nu A_{\lambda cd} + \frac{2}{3} f^{cda}{}_g f^{efgb} A_{\mu ab} A_{\nu cd} A_{\lambda ef}\right) ,
\end{split}
\end{equation}
where the gauge covariant derivative is $D_\mu X_a = \partial_\mu X_a - f^{cdb}{}_a A_{\mu cd} X_b $ and $\Psi$ and $A$ are fermion and gauge fields, respectively.
Note that the gauge field carries two color indices.
BLG theory was shown to be color dual in \cite{Bargheer:2012gv} and the color dual properties of the theory were subsequently investigated in \cite{Huang:2012wr, Huang:2013kca}.

The double copy between BLG theory and generalized ZM theory is reminiscent of the double copy between $\mathcal{N}=4$ SYM and the NLSM in that the resulting theory should involve a photon-matter supermultiplet.\footnote{In the case of SYM $\otimes$ NLSM, the resulting theory is Dirac-Born-Infeld-Volkov-Akulov (DBIVA) theory \cite{Cachazo:2014xea, He:2016vfi, Cachazo:2016njl, Carrasco:2016ldy}.}
Performing the BLG $\otimes$ ZM double copy requires the $D=3$ version of \eqref{eq:DC_rule}, specifically,
\begin{equation}
\label{eq:DC_rule_3D}
f^{a_1 a_2 a_3 a_4} X_1^{a_1} X_2^{a_2} X_3^{a_3} X_4^{a_4} \to \varepsilon^{\mu_1 \mu_2 \mu_3} \partial_{\mu_1} X_1 \partial_{\mu_2} X_2 \partial_{\mu_3} X_3 X_4.
\end{equation}
Implementing the off-shell double copy might appear to be a straightforward application of \eqref{eq:DC_rule_3D} to \eqref{eq:BLGLagrangian}, but the situation proves to be more nuanced.
Instead, we begin by double copying the on-shell four-point amplitude.
The relevant BLG superamplitude is
\begin{equation}
\label{eq:BLG_amp}
\mathcal{A}_4 \propto \delta^3(P) \delta^8(Q) \frac{f^{a_1a_2a_3a_4}}{\langle 12 \rangle \langle 23 \rangle \langle 31 \rangle} ,
\end{equation}
where the notation is familiar from $\mathcal{N}=4$ SYM or can be found in \cite{Huang:2010rn} along with the calculation of the amplitude.
This amplitude expresses two matter particles (scalars or fermions) exchanging a gluon.
Under the on-shell version of \eqref{eq:DC_rule_3D}, the structure constants are mapped to $\varepsilon_{\mu\nu\rho} p_1^\mu p_2^\nu p_3^\rho \propto \langle 12 \rangle \langle 23 \rangle \langle 31 \rangle$, meaning that the four-point amplitude in the product theory is simply
\begin{equation}
\label{eq:BLG_DC_amp}
\mathcal{M}_4 \propto \delta^3(P) \delta^8(Q).
\end{equation}
This amplitude should describe two matter particles exchanging a photon rather than a gluon.

Turning to the off-shell double copy, the fundamental issue is that \eqref{eq:DC_rule_3D} does not describe how the gauge terms in \eqref{eq:BLGLagrangian} should transform.
For example, consider one of the cubic terms $\mathcal{L}_3 \sim f^{abcd} \partial_\mu X_a^I A_{\mu bc} X_d^I$ appearing in \eqref{eq:BLGLagrangian}.
When performing the double copy, color factors are lined up with their associated duality-satisfying numerators.
Although the color factors in $\mathcal{L}_3$ and \eqref{eq:DC_rule_3D} are both $f^{a_1 a_2 a_3 a_4}$, the first is a three-point kinematic object while the second is a four-point kinematic object, making the two completely incompatible.
A different way to phrase the issue is that double copying objects with multiple color indices (like $A_{\mu ab}$) is beyond the purview of \eqref{eq:DC_rule_3D}.
The standard BLG Lagrangian is simply not in a form where \eqref{eq:DC_rule_3D} can be applied.
If the gauge field were integrated out of BLG theory, then it would be trivial to apply \eqref{eq:DC_rule_3D} at the Lagrangian level.
Integrating out the gluon would introduce a tower of non-local interactions, the first of which corresponds to \eqref{eq:BLG_amp}, which we have already double copied.

While BLG theory has maximal $\mathcal{N}=8$ supersymmetry, it is natural to consider double copying generalized ZM theory with less supersymmetric theories like $\mathcal{N}=6$ Aharon-Bergman-Jafferis-Maldacena (ABJM) theory \cite{Aharony:2008ug}.
Just like BLG theory, ABJM theory is color dual with subtle fundamental BCJ relations \cite{Bargheer:2012gv, Huang:2013kca, Sivaramakrishnan:2014bpa}.
ABJM theory can also be formulated in terms of four-index structure constants but with fewer index symmetries concomitant with the reduction in supersymmetry \cite{Bagger:2008se, Gustavsson:2008dy}.
ABJM structure constants need only be antisymmetric in the first two indices and the last two indices separately.
Imposing antisymmetry in all four indices converts the theory back to that of BLG \cite{Bagger:2008se, Gustavsson:2008dy}.
Since the kinematic structure constants of generalized ZM theory are fully antisymmetric we have chosen to focus on BLG theory.
While there does not appear to be any obstruction to double copying ABJM theory with generalized ZM theory on-shell, performing the off-shell double copy would require a version of the ABJM Lagrangian with the gluon integrated out.

\subsection{Modified Chern-Simons theory}
\label{sec:modified_CS_CK}

CS theory is an extraordinarily special quantum field theory; it is integrable, conformal, and topological, with deep connections to knot theory \cite{Witten:1988hf}.
It also happens to be one of the few known off-shell color-dual theories \cite{Ben-Shahar:2021zww}.
This section describes a modification of CS theory that incorporates the $n$-index structure constants.
The role of the generalized Jacobi identity in \eqref{eq:colorrel} will be to ensure gauge invariance of the modified theory.
A potentially unphysical gauge condition is identified that would ensure off-shell color-kinematics duality.

We begin by reviewing a superspace formulation of CS theory.
Following \cite{Axelrod:1991vq} we introduce a superfield
\begin{align}
\Psi^a = c^a + \theta_\mu A^{a\mu} + \theta_\mu \theta_\nu C^{a\mu\nu} + \varepsilon^{\mu\nu\rho} \theta_\mu \theta_\nu \theta_\rho n^a ,
\end{align}
where $\theta^\mu$ is a Grassmann variable, $A^a_\mu$ is the non-abelian gauge field, $c^a$ is one ghost, $C^{a\mu\nu} = \frac{1}{2} \varepsilon^{\mu\nu\rho}\partial_\rho \bar{c}^a$ is the dual of the other ghost, and $n^a$ is a non-dynamical field.
The Lagrangian for CS theory is then
\begin{align}
\label{eq:CSL}
\mathcal{L} = \frac{k}{2\pi} \int d^3 \theta ~ \frac{1}{2} \Psi^a Q \Psi^a + \frac{1}{3!} f^{abc} \Psi^a \Psi^b \Psi^c ,
\end{align}
where $Q=\theta^\mu \partial_\mu$.
This Lagrangian is invariant under the gauge transformation $\Psi^a \to \Psi^a + Q \Omega^a+f^{abc} \Psi^b \Omega^c$ provided that the structure constants obey the Jacobi identity.
Note that the coupling constant in \eqref{eq:CSL} has been set to unity.

A possible modification of \eqref{eq:CSL} is
\begin{align}
\label{eq:generalCS}
\mathcal{L} = \int d^D \theta ~ \frac{1}{2} \Psi^a Q \Psi^a + \frac{1}{D!} f^{a_1 a_2 ..a_D} \Psi^{a_1} \Psi^{a_2}...\Psi^{a_D},
\end{align}
where the rank of the color object is linked to the spacetime dimension.\footnote{Although it is very natural to take the rank of the structure constants, the spacetime dimension, and the number of Grassmann variables to be equal, it can be fruitful to make other choices.  For example, \cite{Ben-Shahar:2024dju, Ben-Shahar:2025dci} generated a color-dual Lagrangian for $BF$ theory by extending the superspace in \eqref{eq:CSL} to four dimensions.}
The coupling constant has been set to unity as before.
Like ordinary CS theory, \eqref{eq:generalCS} makes no reference to a metric, so the theory is topological and classically conformal.
Even though the theory is topological, we are only concerned with perturbative dynamics, so the prefactor of \eqref{eq:generalCS} has been normalized away.
Generally, the coupling constant of \eqref{eq:generalCS} is dimensionful, so classical conformal invariance is likely broken at the quantum level.
Since the structure constants are totally antisymmetric, the superfield needs to be Grassmann odd.
The analog of the gauge transformation is $\Psi^a \to \Psi^a+\delta\Psi^a$, where
\begin{align}
\label{eq:CSgauge}
\delta\Psi^{a_1} =  Q\Omega^{a_1} + \frac{1}{(D-2)!}f^{a_1 a_2 a_3...a_D} \Psi^{a_2} \Psi^{a_3} ... \Psi^{a_{D-1}}\Omega^{a_D}.
\end{align}
In order for $\Psi$ to remain Grassmann odd under a gauge transformation, $\Omega$ must be Grassmann even and $D$ must be odd.\footnote{The standard \emph{abelian} CS term in higher dimensions $A \wedge F \wedge F...$ also requires an odd dimension like \eqref{eq:generalCS}, but otherwise the two theories seem quite different.}
A detailed proof of gauge invariance can be found in Appendix \ref{sec:CSappendix}, but the critical step is to use the color relation in \eqref{eq:colorrel}.
Curiously, much of the proof of gauge invariance would still hold in an even spacetime dimension.

Finding the gauge condition necessary for off-shell color-kinematics duality requires the Feynman rules for the theory.\footnote{Note that proving off-shell color-kinematics duality at tree level in the superspace formulation automatically extends the property to all loop orders because $\Psi$ already incorporates ghosts.}
The propagator and vertex are
\begin{equation}
\frac{b}{p^2} \delta^D (\theta -\theta') \qquad \text{and} \qquad i f^{a_1...a_D} \int d^D\theta ,
\end{equation}
respectively, where $b=ip^\mu \tfrac{\partial}{\partial \theta^\mu} = \partial^\mu \tfrac{\partial}{\partial \theta^\mu}$.
The sum of off-shell numerators appearing in the kinematic Jacobi relation in \eqref{eq:N_Jacobi_general} is
\begin{equation}
\label{eq:psi_kin_jac}
\int d^D \theta ~ b(\Psi_1 ...\Psi_{D-1}) \Psi_D ... \Psi_{2D-2} +\text{sgn cyc}(1,2,...,D) 
\end{equation}
up to an irrelevant overall factor.\footnote{The color factor has been stripped off in \eqref{eq:psi_kin_jac} so the subscript $i$ in $\Psi_i$ is a particle label, not a color index.}
We have verified for $3\leq D \leq 51$, including both even and odd dimensions, that
\begin{equation}
\label{eq:psi_identity}
\begin{split}
&b(\Psi_1 ...\Psi_{D-1}) \Psi_D + \text{sgn cyc}(1,2,...,D)\\
& =\frac{D-1}{2} b(\Psi_1...\Psi_D) + \frac{1}{D-2} \sum \limits_{\sigma \in P} \text{sgn}(\sigma) b(\Psi_{\sigma_1}...\Psi_{\sigma_{D-2}}) \Psi_{\sigma_{D-1}} \Psi_{\sigma_D} .
\end{split}
\end{equation}
The set $P$ in the sum denotes the set of distinct partitions of $\{1,2,...,D\}$ into two subsets $\{\sigma_1, ...\sigma_{D-2} \}$ and $\{\sigma_{D-1}, \sigma_D\}$ where the ordering within the subsets does not matter.
Fixing $\sigma_{D-1}$ and $\sigma_D$ determines the entries of the other subset, so there are $\binom{D}{2}$ terms in $P$.
For example, in five dimensions the partitions into $( \sigma_1 \sigma_2 \sigma_3 \vert \sigma_4 \sigma_5 )$ are
\begin{equation}
\begin{split}
P =\{ & (123 \vert 45), (124 \vert 35), (125 \vert 34), (134 \vert 25), (135 \vert 24), \\
&(145 \vert 23), (234 \vert 15), (235 \vert 14), (245 \vert 13), (345 \vert 12) \} .
\end{split}
\end{equation}
Inserting \eqref{eq:psi_identity} into \eqref{eq:psi_kin_jac} and then integrating by parts on
\begin{equation}
\frac{D-1}{2} \int d^D \theta ~ b(\Psi_1...\Psi_D) \Psi_{D+1} ... \Psi_{2D-2}
\end{equation}
results in a sum of terms, each of which involves $b$ acting on a product of $D-2$ fields.
The off-shell kinematic Jacobi identity \eqref{eq:N_Jacobi_general} would then hold provided that
\begin{equation}
\label{eq:psi_gauge_cond}
b(\Psi_1...\Psi_{D-2}) = 0 .
\end{equation}
In three dimensions this is just the Lorenz gauge condition, $b(\Psi)=0$.
Lorenz gauge alone does not generically imply \eqref{eq:psi_gauge_cond} since $b$ is a second order differential operator.
For dimensions greater than three, \eqref{eq:psi_gauge_cond} becomes non-linear, somewhat like Dirac gauge, $A_\mu A^\mu=\text{constant}$.
Determining the viability of \eqref{eq:psi_gauge_cond} as a gauge condition is beyond the scope of this work.
If it is not a legal gauge condition, then it might be possible to enforce it ad hoc by adding a Lagrange multiplier term to the action.

Beyond gauge-fixing subtleties, the theory in \eqref{eq:generalCS} exhibits additional atypical features that may point to a pathology.
The gauge transformation in \eqref{eq:CSgauge} mixes the fields non-linearly, obscuring whether the gauge redundancy properly absorbs unphysical degrees of freedom.
Furthermore, expanding the superfield in the Lagrangian shows that the gluon sector (or $\mathcal{O}(\theta)$ piece of $\Psi$) kinetically mixes with other sectors, very much unlike CS or YM theory.
Even if \eqref{eq:generalCS} does turn out to be pathological, such theories have proven useful in the study of color-kinematics duality before.
For instance, the semi-abelian Yang-Mills theory presented in \cite{Edison:2023ulf} is not gauge invariant, but it successfully encodes the tree-level dynamics of the NLSM, SDYM theory, and CS theory in a manifestly color-dual fashion.
At the very least, \eqref{eq:generalCS} is a natural extension of normal 3D CS theory that helps motivate the color relations in \eqref{eq:colorrel}.

\subsection{Vector theory}
\label{sec:vector_theory}

We mention one final method for constructing a manifestly color-dual theory respecting the generalized Jacobi identity in \eqref{eq:colorrel}.
As already noted, the Levi-Civita symbol respects \eqref{eq:colorrel}, so a color-dual theory can be obtained by mapping every barred color index of generalized BAS theory \eqref{eq:generalizedBAS} to a spatial index ($\bar{a}_i \to \mu_i$) while taking the kinematic structure constants to be the Levi-Civita symbol.
This produces the theory
\begin{equation}
\label{eq:theoryA}
\mathcal{L} = -\frac{1}{2} \partial_\mu A^a_\nu \partial^\mu A^{a \nu} + \frac{1}{D!} f^{a_1 ... a_D} \varepsilon^{\mu_1...\mu_D} A^{a_1}_{\mu_1}...A^{a_D}_{\mu_D} .
\end{equation}
The Lagrangian does not include ghosts so manifest color-kinematics duality is limited to tree level.
More importantly, the theory is pathological since the Lagrangian is not gauge invariant.
However, as mentioned in the previous subsection, pathological theories have been informative about the double copy in the past.
Like the other theories described in this section, \eqref{eq:theoryA} is classically conformal.

\section{Conclusions and future directions}
\label{sec:conclusions}

Color-kinematics duality and the double copy have proven invaluable tools in the modern $S$-matrix program that enable otherwise intractable calculations and point to profound structure.
The double copy was initially formulated for gauge bosons transforming in the adjoint, where every color object in the theory was constructed out of standard three-index structure constants, $f^{abc}$.
Later, the double copy was extended to include additional color structures such as $f^{abcd}$, $d^{abc}$, and $T^a_{ij}$ \cite{Johansson:2014zca, Johansson:2015oia, Carrasco:2022jxn, Bagger:2007jr, Gustavsson:2007vu, Bargheer:2012gv, Huang:2012wr, Huang:2013kca}.
The goal of this paper was to expand the color structures participating in the double copy to include the infinite family of antisymmetric $n$-index structure constants.

In this work, we introduced the generalized Jacobi identity in \eqref{eq:colorrel} and presented four off-shell color-dual theories incorporating the $n$-index structure constants.
The first theory was the generalization of bi-adjoint scalar theory in \eqref{eq:generalizedBAS}, the second was the generalization of Zakharov-Mikhailov theory in \eqref{eq:generalizedZM}, the third was the modified Chern-Simons theory in \eqref{eq:generalCS} subject to the condition in \eqref{eq:psi_gauge_cond}, and the fourth was the pathological vector theory in \eqref{eq:theoryA}.
All four theories are color-dual off-shell, with all of the theories except the vector theory exhibiting this property at all loop orders.
All of the theories also happen to be classically conformal, with the modified CS theory being trivially topological as well.
The two scalar theories share several additional properties including soft theorems and on-shell recursion relations as well as conserved currents associated with color-kinematics duality.
We discussed the double copy between generalized ZM theory and BLG theory and between generalized ZM theory and itself.
Unlike the two-dimensional case, where the amplitudes of ZM theory and the squared theory are either trivial or subtle, we found that the amplitudes of the $D$-dimensional counterparts are unambiguous and non-zero.

This paper prompts many new directions for future research.
One natural question regarding the color relations in \eqref{eq:colorrel} is whether there are any analogous generalizations for $d^{abc}$ or $T_{ij}^a$.
Some evidence supporting this possibility is given in Appendix \ref{sec:ZM_appendix}, where we present a $d^{abc}$-extension of 2D ZM theory that respects off-shell color-kinematics duality.
Since generalizing ZM theory was central to the $f^{abc...}$ discussion, a $d^{abc}$ extension suggests that there may be additional multi-index symmetric color modifications of the double copy.
More generally, for every property of a known color-dual theory, we may ask whether a similar property holds for an analogous theory involving the $n$-index structure constants.
For example, is there a color-dual analog of YM theory involving $n$-index structure constants?
Could there be any classical or numerical solutions that double copy \cite{Monteiro:2014cda, Luna:2015paa, Cheung:2022mix}?
Is there a notion of color-ordered partial amplitudes?
This is not entirely obvious since $f^{abcd...}$ was not constructed from a trace of generators.
One option may be to use a variant of the Del Duca-Dixon-Maltoni basis \cite{DelDuca:1999rs}.
If a suitable version of color-ordering is found, are there KLT relations \cite{Kawai:1985xq}, fundamental BCJ identities \cite{Bern:2008qj}, or Kleiss-Kuijf relations \cite{Kleiss:1988ne}?
These relations have been found for the $f^{abcd}$ case, but the situation is more complicated than the $f^{abc}$ case \cite{Huang:2013kca}.
If fundamental BCJ relations are found in the $n$-index case, does the theory have hidden zeroes \cite{Bartsch:2024amu, Arkani-Hamed:2023swr, Arkani-Hamed:2024nhp, Arkani-Hamed:2024yvu}?
Is there a CHY description or ambitwistor string for any of the theories involving $f^{abcd...}$ \cite{Cachazo:2013hca, Cachazo:2013iea, Mason:2013sva}?
Finally, the interaction term in the generalized ZM theory computes an oriented volume in momentum space.
Is there a more elegant geometrical interpretation of its scattering amplitudes \cite{Arkani-Hamed:2012zlh, Arkani-Hamed:2013jha}?

\acknowledgments

We would like to thank JJ Carrasco, Alex Edison, and Nic Pavao for helpful discussions on related topics.
We are also grateful to Alex Edison for comments on a previous draft.
J.M. gratefully acknowledges support from Colorado Mesa University's Department of Physical and Environmental Sciences.
Feynman diagrams were typeset using TikZ-Feynman \cite{Ellis:2016jkw}.

\appendix
\section{Five-index structure constants}
\label{sec:f5appendix}

As argued in the main text, $\varepsilon^{abc...}$ should be a valid realization of $f^{abc...}$ for any number of indices.
A natural question is whether there are additional realizations besides the Levi-Civita symbol.
Of course, for the three-index structure constants there are many choices.
Given the generators $T^a$ of some group, the three-index structure constants are constructed from a trace of an antisymmetric sum of generators,
\begin{equation}
f^{a_1 a_2 a_3} \propto \text{Tr}(T^{a_1} T^{a_2} T^{a_3}) + \text{antisym}(a_1,a_2,a_3) .
\end{equation}
Such a construction vanishes for four indices because of the cyclic nature of the sum, and the same holds true for any even number of indices.
Indeed, the essentially unique choice for the four-index structure constants is the Levi-Civita symbol \cite{2007arXiv0712.1398N, Gauntlett:2008uf, Papadopoulos:2008sk}.
The next most elaborate case is the five-index structure constants, which could be assembled from a trace of generators as
\begin{equation}
f^{a_1 a_2 a_3 a_4 a_5} \propto \text{Tr}(T^{a_1} T^{a_2} T^{a_3} T^{a_4} T^{a_5}) + \text{antisym}(a_1,a_2,a_3,a_4,a_5) .
\end{equation}
Taking the Gell-Mann matrices of $SU(3)$ as an example and normalizing $ f^{45678}$ to 1, the non-zero values of these structure constants are
\begin{align}
& f^{12345} = f^{12367} = -\frac{\sqrt{3}}{2} \\
& f^{12458} = f^{23478}= -\frac{1}{2} \\
& f^{12678} = f^{13468} = f^{13578} = f^{23568} = \frac{1}{2} \\
& f^{45678} = 1 ,
\end{align}
with all other non-zero entries obtained through antisymmetrization.
Unsurprisingly, this set of five-index structure constants satisfies \eqref{eq:colorrel}.

\section{Gauge invariance of the modified Chern-Simons theory}
\label{sec:CSappendix}

This appendix shows that the modification of Chern-Simons theory in \eqref{eq:generalCS} is invariant under the gauge transformation in \eqref{eq:CSgauge}.
Under such a transformation the Lagrangian changes by
\begin{align}
& \delta \mathcal{L} = \delta \mathcal{L}_\text{kin} + \delta \mathcal{L}_\text{int}\\
& \delta \mathcal{L}_\text{kin} = \int d^D \theta ~ \frac{1}{2} ( \Psi^a Q \delta \Psi^a + \delta \Psi^a Q \Psi^a )\\
& \delta \mathcal{L}_\text{int} = \int d^D \theta ~ \frac{1}{D!} f^{a_1 a_2...a_D} \sum\limits_{i=1}^{D} \Psi^{a_1} \Psi^{a_2}...\delta \Psi^{a_i}...\Psi^{a_D}.
\end{align}
After integrating by parts (IBP) and anti-commuting $\theta$ past $\Psi$, the variation of the kinetic term becomes
\begin{align}
\delta \mathcal{L}_\text{kin} &= \int d^D \theta ~ Q \Psi^a \delta \Psi^a\\
&= Q \Psi ^a Q \Omega^a + \frac{1}{(D-2)!} f^{a_1 a_2...a_D} Q \Psi^{a_1} \Psi^{a_2}...\Psi^{a_{D-1}} \Omega^{a_D} ,
\end{align}
where the first term vanishes after IBP since $Q$ is nilpotent.
Expanding the variation of the interaction term yields
\begin{align}
& \delta \mathcal{L}_\text{int} = \delta \mathcal{L}_1 + \delta \mathcal{L}_2\\
& \delta \mathcal{L}_1=\int d^D \theta ~ \frac{1}{D!} f^{a_1 a_2...a_D} \sum\limits_{i=1}^{D} \Psi^{a_1} \Psi^{a_2}...Q\Omega^{a_i}...\Psi^{a_D}\\
& \begin{aligned}
\delta \mathcal{L}_2=\int d^D \theta ~ \frac{1}{D! (D-2)!} f^{a_1 a_2...a_D} \sum\limits_{i=1}^{D} \Psi^{a_1} \Psi^{a_2}...\Psi^{a_{i-1}} \qquad\\
\times (f^{a_i b_2...b_D}\Psi^{b_2}...\Psi^{b_{D-1}}\Omega^{b_D})\Psi^{a_{i+1}}...\Psi^{a_D}.
\end{aligned}
\end{align}
After anti-commuting $Q\Omega^{a_i}$ and relabeling indices, $\delta \mathcal{L}_1$ becomes
\begin{align}
\delta \mathcal{L}_1&=\int d^D \theta ~ \frac{1}{(D-1)!} f^{a_1 a_2...a_D} \Psi^{a_1} \Psi^{a_2}...\Psi^{a_{D-1}}Q\Omega^{a_D}\\
&=-\int d^D \theta ~ \frac{1}{(D-1)!} f^{a_1 a_2...a_D} Q(\Psi^{a_1} \Psi^{a_2}...\Psi^{a_{D-1}})\Omega^{a_D}.
\end{align}
This cancels with $\delta \mathcal{L}_\text{kin}$ after distributing the $Q$ and rearranging terms.
The only term remaining in $\delta \mathcal{L}$ is $\delta \mathcal{L}_2$.
The factor of $f^{a_i b_2...b_D}$ can be freely pulled out to the front while $\Omega^{b_D}$ can be sent to the end of the expression.
Anti-commuting $\Psi^{a_{i+1}}$ past all of the $\Psi^{b_j}$ fields introduces a factor of $(-1)^{D-2}$.
Repeating this process a total of $D-i$ times brings all of the $\Psi^{a_k}$ fields to the right of all of the $\Psi^{b_j}$ fields to give
\begin{align}
\begin{split}
\delta \mathcal{L}_2&=\int d^D \theta ~ \frac{1}{D! (D-2)!} \sum\limits_{i=1}^{D} (-1)^{(D-2)(D-i)} f^{a_1 a_2...a_D} f^{a_i b_2...b_D}\\
&\qquad \times \Psi^{a_1} \Psi^{a_2}...\Psi^{a_{i-1}}\Psi^{a_{i+1}}...\Psi^{a_D} \Psi^{b_2}...\Psi^{b_{D-1}} \Omega^{b_D}.
\end{split}
\end{align}
Converting $f^{a_1 a_2...a_D}$ to $f^{a_1 a_2...a_{i-1} a_{i+1} ...a_D a_i}$ introduces a factor of $(-1)^{D-i}$, so that
\begin{align}
\begin{split}
\delta \mathcal{L}_2&=\int d^D \theta ~ \frac{1}{D! (D-2)!} \sum\limits_{i=1}^{D} (-1)^{(D-1)(D-i)} f^{a_1 a_2...a_{i-1} a_{i+1}... a_D a_i} f^{a_i b_2...b_D} \\
&\qquad \times \Psi^{a_1} \Psi^{a_2}...\Psi^{a_{i-1}}\Psi^{a_{i+1}}...\Psi^{a_D} \Psi^{b_2}...\Psi^{b_{D-1}}\Omega^{b_D}.
\end{split}
\end{align}
Now that the indices on the structure constants and superfields are aligned and in ascending order, it is straightforward to relabel indices to obtain
\begin{align}
\label{eq:EvenFunny}
\begin{split}
\delta \mathcal{L}_2&=\int d^D \theta ~ \frac{1}{D! (D-2)!} \sum\limits_{i=1}^{D} (-1)^{(D-1)(D-i)} f^{a_1 a_2... a_{D-1} c} f^{c a_D b_1...b_{D-2}} \\
&\qquad \times \Psi^{a_1} \Psi^{a_2}...\Psi^{a_{D}} \Psi^{b_1}...\Psi^{b_{D-3}}\Omega^{b_{D-2}}
\end{split}
\\
\label{eq:before}
&=\int d^D \theta ~ \frac{f^{a_1 a_2... a_{D-1} c} f^{c a_D b_1...b_{D-2}}}{(D-1)! (D-2)!} \Psi^{a_1} \Psi^{a_2}...\Psi^{a_D} \Psi^{b_1}...\Psi^{b_{D-3}}\Omega^{b_{D-2}} ,
\end{align}
where in the second line we have used that $D$ is odd so that $(-1)^{(D-1)(D-i)}=1$.
Since $a_1$, $a_2$,...,$a_D$ are dummy indices, we may replace the following factor of \eqref{eq:before} with a cyclic sum over the dummy indices
\begin{align}
&f^{a_1 a_2... a_{D-1} c} f^{c a_D b_1...b_{D-2}} \Psi^{a_1} \Psi^{a_2}...\Psi^{a_D} \\
&\label{eq:VanishByJacobi1}
= \frac{1}{D}\left[ f^{a_1 a_2... a_{D-1} c} f^{c a_D b_1...b_{D-2}} \Psi^{a_1} \Psi^{a_2}...\Psi^{a_D} + \text{cyc}(a_1, a_2,...,a_D)\right].
\end{align}
Each term in the sum contains a cyclic rotation of the superfields.
Anti-commuting the superfields past each other to bring them to canonical ordering introduces no net minus signs because $D$ is odd.
Once all of the superfields are in the same order, they can be factored out of the sum to leave
\begin{align}
\label{eq:VanishByJacobi2}
\frac{1}{D}\left[ f^{a_1 a_2... a_{D-1} c} f^{c a_D b_1...b_{D-2}}+ \text{cyc}(a_1, a_2,...,a_D)\right] \Psi^{a_1} \Psi^{a_2}...\Psi^{a_D}
\end{align}
which vanishes by the Jacobi identity in \eqref{eq:colorrel}.
Thus, $\delta \mathcal{L}_2$ vanishes and \eqref{eq:generalCS} is invariant under a gauge transformation.

In the main text we argued that $D$ must be odd in order for $\Psi$ to remain Grassmann odd under a gauge transformation.
Curiously, if $\Psi$ somehow remained anti-commuting in an \emph{even} spacetime dimension, then many of the preceding manipulations would remain unaffected.
For example, when $D$ is even then the sum over $(-1)^{(D-1)(D-i)}$ in \eqref{eq:EvenFunny} still vanishes.
Even if the sum in \eqref{eq:EvenFunny} did not vanish, anti-commuting all of the superfields in \eqref{eq:VanishByJacobi1} past each other would convert \eqref{eq:VanishByJacobi2} into a signed cyclic sum and \eqref{eq:VanishByJacobi2} would still vanish by \eqref{eq:colorrel}.
It seems then that the requirement of an odd spacetime dimension stems entirely from the need for $\Psi$ to transform homogeneously as Grassmann odd under a gauge transformation.

\section{Color-dual Zakharov-Mikhailov theory with $d^{abc}$}
\label{sec:ZM_appendix}

This paper has been chiefly concerned with \emph{antisymmetric} structure constants $f^{abc} =-2i\text{Tr}([T^a, T^b] T^c)$, but the double copy framework also admits \emph{symmetric} color structures $d^{abc}=2\text{Tr}(\{T^a ,T^b\}T^c)$ \cite{Carrasco:2022jxn}.\footnote{In our conventions $\text{Tr}(T^a T^b) = \frac{1}{2} \delta^{ab}$.}
In this appendix, we will show that 2D ZM theory can be augmented with a symmetric color structure while maintaining off-shell color-kinematics duality.
The simplest modification is
\begin{equation}
\label{eq:ZM_dabc}
\mathcal{L} = -\frac{1}{2} \partial_\mu \phi^a \partial^\mu \phi^a + g_1 f^{abc} \phi^a \varepsilon^{\mu\nu} \partial_\mu \phi^b \partial_\nu \phi^c + \frac{g_2}{\alpha'} d^{abc} \phi^a \phi^b \phi^c ,
\end{equation}
where $\alpha'$ has units of length squared.
Ordinary 2D ZM theory is recovered by setting $g_2=0$ in \eqref{eq:ZM_dabc}.
To demonstrate off-shell color-kinematics duality, we must first enumerate the algebraic relations between the different color structures.
We restrict our attention to $U(N)$ since the color relations for $SU(N)$ involve $1/N$ corrections.
Aside from $f^{abc}$ being totally antisymmetric and $d^{abc}$ being totally symmetric,\footnote{The kinematic vertices presented in this appendix all trivially possess the correct symmetry properties.} the two color objects obey three central identities,
\begin{align}
& \label{eq:ff_cyc} f^{a_1 a_2 c} f^{c a_3 a_4} + \text{cyc}(1,2,3) = 0\\
& \label{eq:df_cyc} d^{a_1 a_2 c} f^{c a_3 a_4} + \text{cyc}(1,2,3) = 0\\
& \label{eq:ff_recast} f^{a_1 a_2 c} f^{c a_3 a_4} = d^{a_1 a_3 c} d^{c a_2 a_4} - d^{a_1 a_4 c} d^{c a_2 a_3} .
\end{align}
The first equation is the standard Jacobi identity.
The second equation will be the most important novel ingredient.
The third equation is somewhat different from the first two in that it allows $ff$-type numerators to be recast as $dd$-type numerators, that is, it provides a way to change color bases.\footnote{Equation \eqref{eq:ff_recast} allows kinematic numerators to be moved between color factors.  This is reminiscent of shuffling kinematic terms between different color factors after blowing up a four-point vertex in YM theory.}
So long as there exists a prescription for unambiguously assigning kinematic numerators to color factors, there does not seem to be a need to enforce \eqref{eq:ff_recast}.
Since the Feynman rules generated by a Lagrangian provide just such a prescription, \eqref{eq:ff_recast} may be set aside temporarily.

Demonstrating off-shell color-kinematics duality for \eqref{eq:ZM_dabc} entails proving the kinematic analogues of \eqref{eq:ff_cyc} and \eqref{eq:df_cyc}.
Up to irrelevant prefactors, the kinematic version of \eqref{eq:ff_cyc} is
\begin{equation}
\varepsilon( p_1, p_2 ) \varepsilon( p_3, p_4 ) + \text{cyc}(1,2,3) = 0 ,
\end{equation}
which was proved as part of \eqref{eq:KinJacIdTemp} and is the reason that ordinary ZM theory is color-dual off-shell \cite{Cheung:2022mix}.
The kinematic version of \eqref{eq:df_cyc} is
\begin{equation}
\varepsilon( p_3, p_4 ) + \text{cyc}(1,2,3) = 0 ,
\end{equation}
which vanishes after imposing off-shell momentum conservation.

There is also a Moyal deformation of ZM theory that is color-dual off-shell,
\begin{equation}
\mathcal{L} = - \frac{1}{2} \partial_\mu \phi^a \partial^\mu \phi^a + \frac{1}{\alpha'} f^{abc} \phi^a \sin[\alpha' \varepsilon( \partial_{\phi^b}, \partial_{\phi^c} )] \phi^b \phi^c ,
\end{equation}
where the subscript on the derivative indicates which field it acts upon \cite{Moyal:1949sk, Cheung:2022mix}.
An interaction with a symmetric color structure can be added as follows,
\begin{equation}
\mathcal{L} = - \frac{1}{2} \partial_\mu \phi^a \partial^\mu \phi^a + \frac{g_1}{\alpha'} f^{abc} \phi^a \sin[\alpha'\varepsilon( \partial_{\phi^b}, \partial_{\phi^c} )] \phi^b \phi^c + \frac{g_2}{\alpha'} d^{abc} \phi^a \cos[\alpha'\varepsilon( \partial_{\phi^b}, \partial_{\phi^c} )] \phi^b \phi^c .
\end{equation}
The kinematic versions of \eqref{eq:ff_cyc} and \eqref{eq:df_cyc} are
\begin{align}
& \sin[\alpha' \varepsilon( p_1, p_2 )] \sin[\alpha' \varepsilon( p_3, p_4 )] + \text{cyc}(1,2,3) = 0\\
& \cos[\alpha' \varepsilon( p_1, p_2 )] \sin[\alpha' \varepsilon( p_3, p_4 )] + \text{cyc}(1,2,3) = 0 ,
\end{align}
which hold off-shell.
Interestingly, enforcing the kinematic version of \eqref{eq:ff_recast} sets $g_1 = g_2$.
The resulting theory is particularly simple,
\begin{equation}
\mathcal{L} = \text{Tr} \left( \partial_\mu \phi \partial^\mu \phi + \frac{1}{\alpha'} \phi \phi e^{-i \alpha' \varepsilon( \stackrel{\leftarrow}{\partial}, \stackrel{\rightarrow}{\partial} )} \phi \right) ,
\end{equation}
where the left/right arrow over the derivative indicates that it acts on the field immediately to the left/right of the exponential.
This Lagrangian makes it appear as though individual generators have been mapped from color space to kinematic space, which is not entirely surprising since $U(N)$ is isomorphic to the group of area preserving diffeomorphisms under the appropriate conditions \cite{Hoppe:1988gk}.

This appendix has focused on off-shell color-dual theories that include a three-index symmetric color object, but it seems plausible that there might be an $n$-index generalization that obeys relations like \eqref{eq:colorrel}.
We leave this to future work.

\bibliographystyle{JHEP}
\bibliography{biblio}

\end{document}